\documentclass[12pt]{article}
\usepackage{amssymb}
\usepackage{fonttable}
\usepackage {xcolor}
\usepackage{braket}

\usepackage{amsmath,amssymb}
\usepackage{graphicx}
\usepackage{braket}
\usepackage{epstopdf}
\title{Impact of Coulomb Correction Factor on Rate of Change of Lepton Fraction during Presupernova Evolution}
\author{
	Asim Ullah$^{1,*}$\thanks{Corresponding author email: asimullah844@gmail.com} 
	\and 
	Jameel-Un Nabi$^{1,2}$
}
\date{
	$^{1}$Faculty of Engineering Sciences, Ghulam Ishaq Khan Institute of Engineering Sciences and Technology, Topi 23640, KP, Pakistan \\
	$^{2}$University of Wah, Quaid Avenue, Wah Cantt 47040, Punjab, Pakistan
}

\begin{document}
\maketitle
\begin{abstract}
We re-examine the weak interaction nuclei having largest contribution to the lepton-to-baryon fraction  (${Y}_e$) by coupling the  stellar weak rates and mass abundances for  post silicon burning conditions during the presupernova evolution of massive stars. The stellar weak rates were  recently calculated by Nabi et al. (2021) employing the fully microscopic pn-QRPA model without invoking the  Brink-Axel hypothesis. We compute the mass abundances for a total of 728 nuclei,  with A = 1--100, using Saha's equation and assuming nuclear statistical equilibrium with the incorporation of Coulomb correction factor to the chemical potential.  We compile a list of top 50 electron capture (\textit{ec}) and $\beta$-decay (\textit{bd}) nuclei, on the basis of largest contribution to $\dot{Y}_e$ for post silicon burning conditions, where 11\% \textit{ec} and 6\% \textit{bd} nuclei debuted due to Coulomb corrections. The calculated mass abundances and corresponding $\dot{Y}_e$ values are enhanced up to 3 orders of magnitude for heavier nuclei once Coulomb corrections were incorporated. This enhancement led to an increment in total $\dot{Y}^{bd}_e$ and $\dot{Y}^{ec}_e$ values, at $Y_e$ = 0.425 ($\rho$ = 2.20$\times$10$^9$ g/cm$^3$), of 80\% and 91\%, respectively. After incorporating the Coulomb corrections, we propose a revised interval of $Y_e$ = 0.423-0.455, where \textit{bd} rates surpass the competing \textit{ec} rates, and is 3.2 \% bigger than the one suggested by Nabi et al. (2021).  
\end{abstract}
\section*{Keywords}
Nuclear Statistical Equilibrium, Electron Capture, Chemical Potential, Lepton-to-baryon Fraction, pn-QRPA, Coulomb Corrections, $\beta$-Decay, Mass Abundances, Presupernova

\section{Introduction}

Core-collapse supernovae are considered to be one of the main contributors to the
formation of heavier elements in the stellar matter. Researchers continue their quest for a better understanding of the dynamics of the core-collapse. The final outcome of the explosion is sensitive to various physical inputs at the start of each stage of the entire process (i.e., collapse, shock formation, and shock propagation). Consequently, the computation of the presupernova stellar structure with the most reliable physical data currently available is highly desirable. \\
The weak interaction plays a key role in
the dynamics of the core collapse. The $\beta$-decay (\textit{bd}) and electron capture (\textit{ec}) rates are amongst the key nuclear-physics inputs that determine both the lepton-to-baryon fraction ($Y_e$) and the entropy at the presupernova phase \cite{Arn96}. They are believed to have several vital consequences during the later phases of stellar evolution of the massive stars. The \textit{ec} effectively reduces the number of electrons available for pressure support, while \textit{bd} has a converse action. Both these processes have a direct influence on the overall $Y_e$ ratio of the stellar core. The neutrinos and antineutrinos, produced during these nuclear weak reactions, escape from the star (during the early phases of presupernova evolution) and thereby channel out the entropy and energy away from the core. These weak interaction rates are important, not only in the accurate determination of the structure of the stellar core, but also play a vital role in the elemental abundance calculation and nucleosynthesis. Weak interactions, in massive stars prior to supernova, are believed to be governed by allowed Fermi and Gamow-Teller (GT) transitions \cite{Beth79}.    \\
Unfortunately, the nuclei which causes largest change to $Y_e$ are neither the most 
abundant ones nor the ones having strongest rates but a combination of the two. The most abundant nuclei usually have high binding energies and consequently smaller weak rates. Similarly the most reactive nuclei tend to be present in small quantities. Thus, in the given scenario,  both the weak rates and abundances must be combined together to determine the most important nuclei. The mass abundances, for a given density ($\rho$), temperature ($T$), and $Y_e$, can be computed from the well known Saha’s equation assuming nuclear statistical equilibrium (NSE) \cite{Har85}.\\
In the past several attempts were made to calculate weak interaction rates and mass abundances in stellar environments in order to gain a better knowledge of the stellar dynamics. Stellar \textit{bd} rates were computed by Takahashi et al. \cite{Tak78} employing the so-called gross theory of beta decay. The drawback in this theory was that the nuclear structure details of the individual nuclei were not taken into account. Instead only a statistical description of the $\beta$-strength function was assumed. Fuller, Fowler, and Newman (FFN) \cite{Ful80,Ful82,Ful82a,Ful85} performed the pioneering calculation by tabulating the weak interaction rates for a total of 226 nuclei with masses in the range 60 $\ge$ A $\ge$ 21, employing the independent-particle model (IPM) with the help of the then available experimental data for astrophysical applications. Their rates resulted in a sizable decrease in the $Y_e$ fraction throughout the core, allowing researchers to gain insight of the development of stars before supernova \cite{Wea85}. In 1994, Aufderheide and colleagues \cite{Auf94} used the IPM model to examine the impact of weak interaction rates in the evolutionary phases of stars following silicon burning and provide a list of the most relevant nuclei in the $Y_e$ range of 0.40—0.50. They expanded on the FFN study by including heavy nuclei (A $>$ 60) and quenched the GT strength explicitly. Later few researchers e.g. \cite{Vet89,El-k94,Wil95} highlighted a flaw in the systematic parameterization used by FFN and Aufderheide et al. Subsequently, the proton-neutron quasiparticle random phase approximation (pn-QRPA) \cite{Nab99,Nab04} and shell model \cite{Lan00} computed stellar weak rates reported that the GT centroids for several important nuclei were not correctly placed by FFN and consequently resulted in a discrepancy between experimental and theoretical data. Heger et. al. \cite{Heg01} used the large scale shell model (LSSM) data to run simulations during the presupernova evolution, assuming stellar rates with mass number A = 45--60 \cite{Lan01}. More recently, Nabi and collaborators \cite{Nab21} performed a study of the presupernova evolution employing the pn-QRPA computed rates. The pn-QRPA model was employed to compute the stellar weak rates and mass abundances of a total 728 nuclei with mass range A = 1–-100. The authors  compiled a list of 50 most important \textit{bd} and \textit{ec} nuclei that had the greatest impact on $Y_e$. The authors further compared their results with the previous calculations of Gross Theory \cite{Fer14} and IPM \cite{Auf94}. \\
In this project, we re-examine the most important weak interaction nuclei during the presupernova evolution by coupling the QRPA computed rates \cite{Nab21} with the newly calculated mass abundances. The pn-QRPA model is particularly suited for stellar rate computations because, unlike conventional stellar rate calculations, it does not use the Brink-Axel hypothesis \cite{Bri58} to compute excited state GT strength distributions. For nearly a thousand nuclei, a good comparison was made between the pn-QRPA model and experimental data (see Figures 11–17 and Tables E–M of \cite{Nab04}). The pn-QRPA model can make use of a large model space (up to 7$\hbar$$\omega$) which permits the calculation of excited states GT strength distribution functions for heavier nuclei. 
The mass abundances, for a total of 728 nuclei with $A$ = 1--100, were calculated by applying Coulomb corrections to nuclear species' chemical potential. Non-ideal effects, particularly Coulomb corrections, have been integrated into the equation of state (EOS) of stellar evolutionary codes since Salpeter's pioneering work on corrections to an ideal plasma at zero temperature \cite{Sal61}. Once Coulomb corrections are included in the EOS, masses of iron core in heavy mass stars, nearing the end of their lives,  were reduced by 5-10\% compared to calculation without corrections \cite{Woo88}. Several authors \cite{Hil84,Nom88,Nom84} have investigated the impact of Coulomb corrections on heat capacity and pressure under conditions suited for studying the evolution of presupernova and supernova explosions in massive stars. The authors in Ref.~\cite{Moch86} emphasized the significance of incorporating the Coulomb corrections to the chemical potential of nuclei in NSE in order to accurately calculate their relative abundances. Hix \& Thielemann \cite{Hix96} later included Coulomb corrections in their Si-burning network calculations. The Coulomb interactions of the ions and electrons in a multicomponent plasma (MCP), such as that produced by NSE, result in a decrease in free energy. This changes the chemical potential of species and, therefore, their abundance in NSE. These corrections to the free energy are significant enough to be included in mass abundance calculations \cite{Nad05}. The Coulomb corrections affect the rate of change of Y$_e$ in two different ways: the abundance distribution of nuclei is changed as discussed above and the \textit{ec} rate for each nucleus is also altered. The change in \textit{ec} rate for a given nucleus is caused by the shift in the threshold energy of capture, as the chemical potential of the nucleus is modified due to corrections. However, the impact on the overall neutronization  rate is not expected to be large since the shift in the threshold energy of capture is of the order 0.3--0.5 MeV \cite{Cou74}. Moreover, the energy of the captured electron is affected by the presence of	the background electron gas. Its energy will be reduced compared to the unscreened case. This will make it easier for a charged projectile nucleus to penetrate the Coulomb barrier of a charged target nucleus thereby enhancing the reaction rate. The magnitude of this effect can be determined using linear response theory \cite{Ito02}. Itoh et al. \cite{Ito02} reported the calculation of screening corrections to \textit{ec} rates via the linear response theory. They concluded that the screening correction to \textit{ec} at high densities were only a few percent. A detailed study of the dynamical screening effects may be found in Ref.~\cite{Hwa21}. There the authors explore the dynamical screening effects on big bang nucleosynthesis (BBN) using a test charge method and concluded that the dynamical screening effects on the primordial abundances were negligible since the thermonuclear rates were hardly changed. NSE computations with Coulomb corrections have recently been  performed by Bravo and García-Senz \cite{Bra99}. Coulomb corrections to the ion ideal gas EOS of matter in NSE were reported to be significant at temperatures $T$ $\leq$ 5-10 GK and densities $\rho$ $\ge$ $10^8$ g/cm$^3$. The neutronization rate increased by roughly 28\% when Coulomb corrections were applied at $T$ = 8.5 GK and $\rho$ = 8$\times$$10^9$ g/cm$^3$  \cite{Bra99}. In this study, we follow the method of Bravo and García-Senz to incorporate the Coulomb corrections in our calculation. We finally compile a revised list of important weak interaction nuclei during presupernova evolutionary phases of massive stars after comparing our results with the recent calculation of Nabi and collaborators \cite{Nab21}. 

The paper is structured as follows. Section~2 provides a brief overview of the formalism used in our calculation. Section~3 discusses and compares our findings with previous calculations. The final section contains summary and concluding remarks.

\section{Formalism}

As mentioned earlier, we used the pn-QRPA computed rates of Nabi and collaborators \cite{Nab21}. For the sake of completeness we reproduce few important equations here. \\
The weak decay rate from the $\mathit{m}$th state (parent nucleus) to
the $\mathit{n}$th state (daughter nucleus) in stellar environment may be determined using the formula given below

\begin{eqnarray}\label{rate}
	\lambda^{ec(bd)}_{mn} =\frac{ln2}{D}[{\phi_{mn}^{ec(bd)}(\rho, T, E_{f})}] \times \left[B(F)_{mn}+\frac{B(GT)_{mn}}{(g_{A}/g_{V})^{-2}} \right].
\end{eqnarray}
The value of constant D appearing in Eq.~(\ref{rate}) was taken as 6143 $s$ \cite{Har09}. The value of $g_{A}$/$g_{v}$, representing the ratio of axial and vector coupling constant, was taken as -1.2694 \cite{Nak10}. It is to be noted that we did not incorporate any explicit quenching factor in our computation. $B(F)_{mn}$ and $B(GT)_{mn}$ are the total reduced transition probabilities due to Fermi and GT interactions, respectively. 

Like any other nuclear model, the pn-QRPA model has limitations and associated uncertainties. The first two factors, stated in Eq.~(\ref{rate}), may be calculated with desirable accuracy depending on the power of the computing machines. The pn-QRPA model is used to calculate the reduced transition probabilities for GT transitions. The biggest source of uncertainty in the pn-QRPA model is associated with the calculation of GT strength functions and computation of parent and daughter energy levels. 
The uncertainty in the Q-values, one of the pn-QRPA model parameters,  is the limiting factor for the calculation of weak interaction rates. The uncertainties involved in the calculations of stellar weak rates can be considerably significant and depend, to a large extent, on the input masses for the calculation of weak rates. It was shown in Ref.~\cite{Hir93} that the calculated half-life of $^{221}$U changed by orders of magnitude when the Q-values changed by less than a couple of MeV’s. This unreasonably large change in calculated half-lives was not an effect of the nuclear structure calculation but due to the reported changes in calculated Q-values.
It is worth mentioning that stellar weak rates are sensitive functions of the difference of parent and daughter energies and by a mere addition of around 0.5 MeV (a typical uncertainty in the calculation of energy levels within the pn-QRPA model) small \textit{bd} rates (close to the value of 10$^{-100}$ s$^{-1}$) may change by orders of magnitude and need to be considered as a limiting factor of the pn-QRPA calculation.

\begin{equation}
	B(F)_{mn} = [{2J_{m}+1}]^{-1}  \mid<n \parallel \sum_{k}t_{\pm}^{k}
	\parallel m> \mid ^{2}
\end{equation}

\begin{equation}\label{bgt}
	B(GT)_{mn} = [{2J_{m}+1}]^{-1}  \mid <n
	\parallel \sum_{k}t_{\pm}^{k}\vec{\sigma}^{k} \parallel m> \mid ^{2},
\end{equation}
where $J_m$, $t_\pm$ and $\vec{\sigma}$ represent the total spin of parent nucleus, isospin operators (raising/lowering) and Pauli spin matrices, respectively.
The phase space integrals $(\phi_{mn})$ for \textit{ec} and \textit{bd} are (hereafter natural units are used, $\hbar=c=m_{e}=1$)
\begin{equation}\label{pc}
	\phi^{ec}_{mn} = \int_{w_{l}}^{\infty} w (w_{m}+w)^{2}({w^{2}-1})^{\frac{1}{2}} F(+Z,w) G_{-} dw,
\end{equation}

\begin{equation}\label{ps}
	\phi^{bd}_{mn} = \int_{1}^{w_{m}} w (w_{m}-w)^{2}({w^{2}-1})^{\frac{1}{2}} F(+Z,w)
	(1-G_{-}) dw,
\end{equation}
In Eqs.~(\ref{pc}) and (\ref{ps}) $w$, $w_l$ and $w_{m}$ denote the total energy (kinetic+rest mass) of the electron, total threshold energy for \textit{ec} and the total energy of \textit{bd}, respectively. The Fermi functions ($F(+ Z, w)$) were calculated employing the procedure used by Gove and Martin ~\cite{Gov71}. $G_{-}$ represents the electron distribution function.
The total \textit{ec} and \textit{bd} rates were calculated using
\begin{equation}\label{ecbd}
	\lambda^{ec(bd)} =\sum _{mn}P_{m} \lambda^{ec(bd)} _{mn}.
\end{equation}
The summation was applied over all the initial and final states and satisfactory convergence in \textit{ec} and \textit{bd} rates was achieved. $P_{m} $ in  Eq.~(\ref{ecbd}) denotes the occupation probability of parent excited states and follows the normal Boltzmann distribution.

Once the NSE is achieved in stellar matter post silicon burning the nuclei are produced and destroyed at the same rate, and the chemical potentials of nucleus \textit{i} and nucleons satisfy the condition
\begin{equation}\label{mu}
	\mu_i = Z_i\mu_p + (A_i-Z_i)\mu_n,
\end{equation}
where $\mu_p$ and $\mu_n$ are the chemical potentials of proton and neutron, respectively. Incorporating Coulomb corrections, the chemical potential of nucleus \textit{i} may be given by
\begin{equation}\label{muC}
	\mu_i = \mu_{i,0} + \mu_{i,C},
\end{equation}
$\mu_{i,0}$ being the chemical potential in the absence of Coulomb effects whereas $\mu_{i,C}$ is the contribution due to the Coulomb corrections. Combining Eq.~(\ref{mu}) \& Eq.~(\ref{muC}) and using the expression for number density $n_i$ (with $n_i$ = $\frac{\rho X_i}{m_i}$), one can find the mass abundance of nucleus \textit{i} as;
\begin{eqnarray}\label{x1}
	X_i(A,Z)= {G_i(Z,A,T)}(2^{-A}) \left({\rho\lambda_T^{3} N_{A} }\right)^{A-1} \nonumber\\
	\times A^{{5}/{2}}x_{p}^{Z}{x_{n}^{A-Z}}\exp
	\left[\frac{-Q_i(A,Z)}{k_BT} \right],
\end{eqnarray}
where the terms $\lambda_T$ (=$\sqrt{\frac{h^2}{2\pi m_Hk_BT}}$) and $N_A$ stands for the Avogadro's number and thermal wavelength, respectively. The mass abundance of free proton ($x_p$) and free neutron ($x_n$) can be found by imposing mass and charge conservations, respectively, and are given by 
\begin{equation}
	\sum_{i} X_i = 1,
\end{equation} 

\begin{equation}
	\sum_{i} \frac{Z_i}{A_i}X_i = Y_e = \frac{1-\eta}{2},
\end{equation}
where $\eta$ stands for neuron excess.  In Eq.~(\ref{x1}), the $Q_i$ denotes the difference in chemical potential between nucleus \textit{i} and its nucleons. This $Q_i$, in the absence of Coulomb factor, is given by
\begin{equation}
	Q_{i,o} = [m_i -Z_im_p-(A_i-Z_i)m_n]c^2,
\end{equation}
Once the Coulomb corrections are incorporated, $Q_i$ becomes
\begin{equation}
	Q_{i} = [m_i -Z_im_p-(A_i-Z_i)m_n]c^2 +\mu_{i,C}-Z_i\mu_{p,C}.
\end{equation} 
The $G_i$(Z,A,T), appearing in Eq.~(\ref{x1}), is the  nuclear partition function (NPF) of the $i^{th}$ nucleus. We use a new recipe, introduced by Nabi and collaborators \cite{Nab16a, Nab16b}, to compute the NPFs, where the states up to excitation energy of 10 MeV are considered as \textit{discrete}. A simple level density function was assumed beyond 10 MeV, and integration was carried out up to a maximum excitation energy of 25 MeV. The missing measured energy levels, if any, were manually inserted into the calculations along with their spins. We refer to \cite{Nab16a, Nab16b} for more information on the computation of NPFs.\\
The time rate of change of $Y_e$ is an important parameter to be monitored throughout the evolutionary stages of massive star prior to supernova.  For nucleus \textit{i}, it is given by
\begin{equation}
	\dot{Y}_{e(i)}^{ec(bd)} = -(+)\frac{X_i}{A_i}\lambda_i^{ec(bd)},
\end{equation}
where the negative (positive) sign is for \textit{ec} (\textit{bd}), $A$ is the mass number and $\lambda^{ec(bd)}$ was computed using Eq.~(\ref{ecbd}).\\
To incorporate the Coulomb correction factor, we follow the procedure of Bravo and García-Senz \cite{Bra99}. In NSE the stellar core is composed of a multicomponent plasma (MCP). Coulomb corrections to the EOS, in such cases, are typically calculated using the additive approximation, where the free energy is computed as the sum of individual free energies of each species. 
Coulomb chemical potential of $i^{th}$ nucleus, using additive approximation, is given by \cite{Yak89}
\begin{equation}
	\mu_{i,C}=k_BTf_C(\Gamma_i),
\end{equation}
where $f_C$ is the Coulomb free energy per ion in units of $k_BT$ and $\Gamma_i$
represents the ion-coupling parameter given by
\begin{equation}
	\Gamma_i=\Gamma_eZ_i^{5/3}=\frac{Z_i^{5/3}e^2}{a_eK_BT},
\end{equation}
with the electron cloud radius $a_e$ given in terms of the electron number density
$n_e$ as
\begin{equation}
	a_e=\left(\frac{3}{4\pi n_e}\right)^{1/3}.
\end{equation}
The relation for the Coulomb corrections to the free energy for strong coupling regime, $\Gamma_i$ $>$ 1 is given by \cite{Oga87}:
\begin{equation}
	f_C (\Gamma_i)=a\Gamma_i+4(b\Gamma_i^{1/4}-c\Gamma_i^{-1/4})+d \ln\Gamma_i-o,
\end{equation}
where the values of the parameters $a, b, c, d$ and $o$ are given by
Ogata \& Ichimaru \cite{Oga87}.\\
For the weak coupling regime $\Gamma_i$ $<$ 1, the corrections may be computed via the expression proposed by Yakoblev \& Shalybkov \cite{Yak89}
\begin{equation}
	f_C (\Gamma_i)= -\frac{1}{\sqrt{3}}\Gamma_i^{3/2}+\frac{\beta}{\gamma}\Gamma_i^{\gamma}.
\end{equation}
It is to be noted that our calculation includes the corrections \textit{only} for non-ideal plasma. There exist different types of fits to interpolate between strong coupling ($\Gamma_i$ $>$ 1) and weak coupling  ($\Gamma_i$ $<$ 1) regimes. For example, Hansen et al. \cite{Han77} introduced a four parameter fit for f($\Gamma$) to interpolate between the two regimes. Yakovlev \cite{Yak89} provided a fit with a different functional form which produced results with less than 4\% deviation from the fit of Hansen et al. \cite{Han77}. Chabrier et al. \cite{Cha98} used the fit of Hansen et al. with three parameters, and their results deviated not more than 1\% from Hansen et al. \cite{Han77}. Later on, Potekhin \& Chabrier \cite{Pot2000} introduced a seven parameter fit that deviate the fit of Chabrier et al. \cite{Cha98} by less than 1 \%.
 
\section{Results and Discussion}

\begin{figure}[h!]
	\centering
	\includegraphics[width=1.\textwidth]{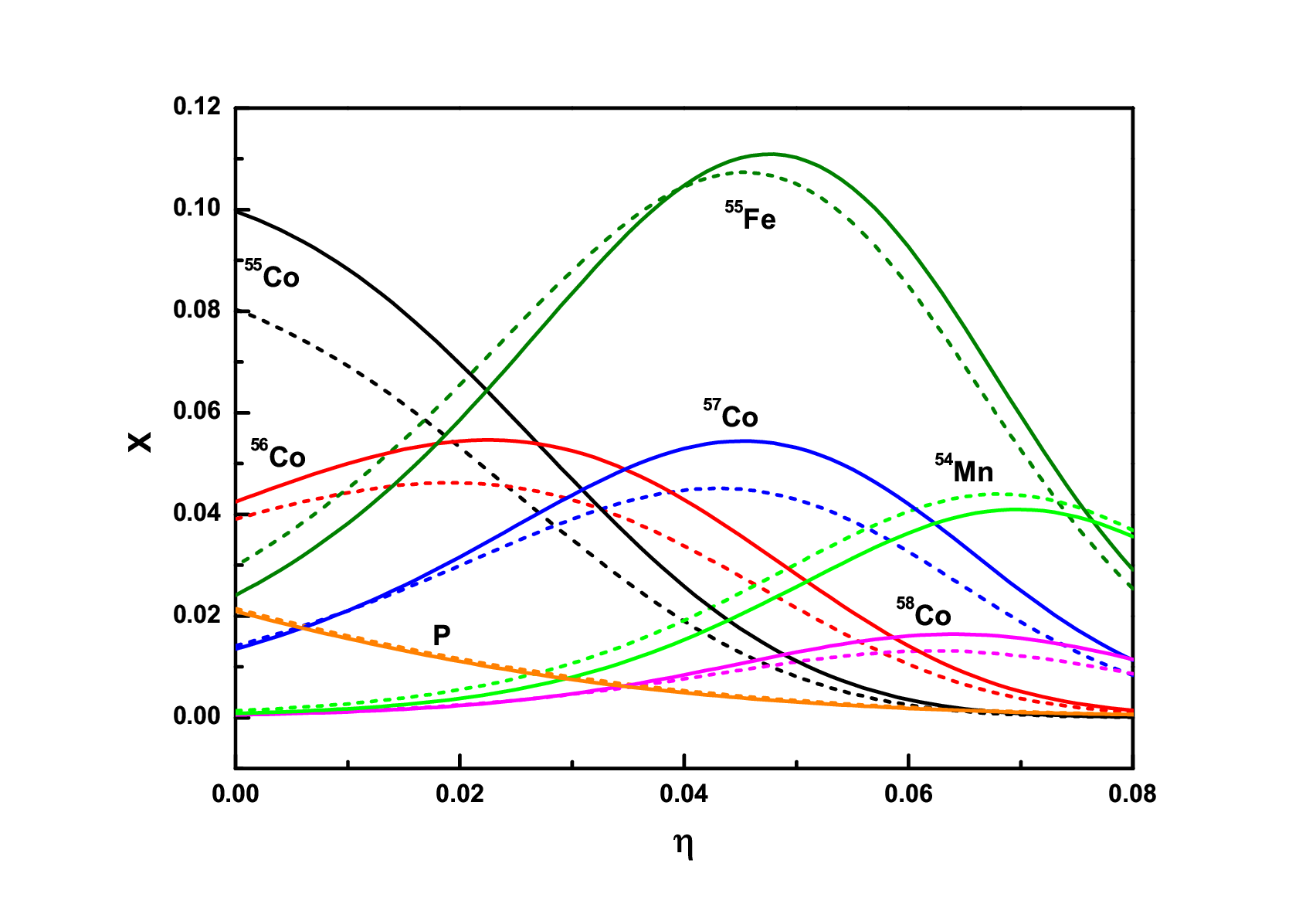}
	\vspace{-8mm}
	\caption{Mass abundances of selected nuclei as a function of neutron excess at $T$ = 8$\times$10$^9$ K and $\rho$ = 2$\times$10$^9$ g/cm$^3$, with (solid lines) and without Coulomb corrections (dashed lines) to the chemical potentials. }
	\label{F1.mf}
\end{figure}

\begin{figure}[h]
	\centering
	\includegraphics[width=1.\textwidth]{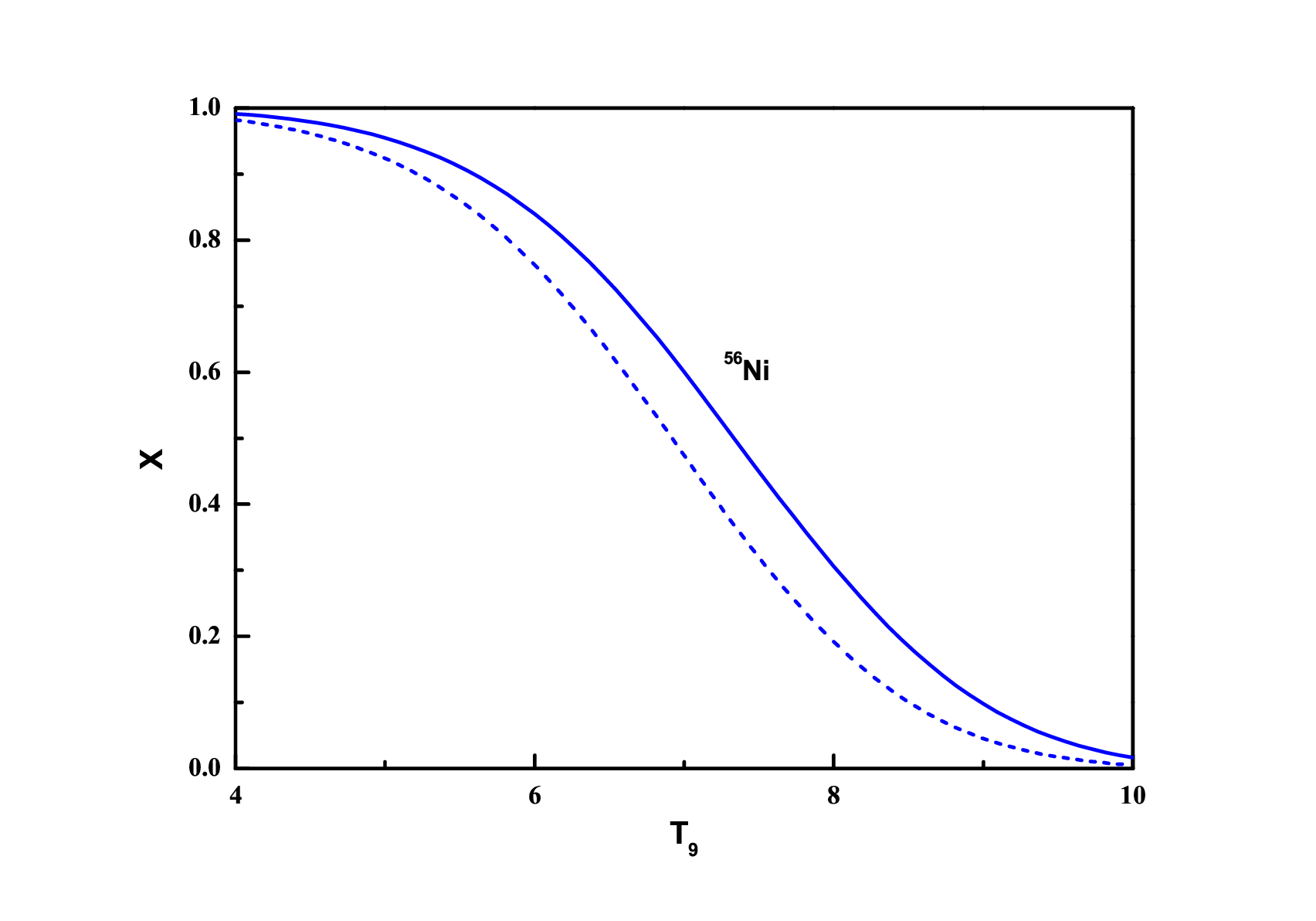}
	\vspace{-8mm}
	\caption{The computed mass abundance of $^{56}$Ni, with  (solid curve) and without Coulomb corrections (dashed curves) to the chemical potentials, as a function of $T_9$ (core temperature in units of 10$^{9}$ K) at fixed $\rho$ = 5$\times$10$^9$ g/cm$^3$ and $Y_e$ = 0.5. }
	\label{Ni56}
\end{figure}

\begin{figure}[h]
	\centering
	\includegraphics[width=1.\textwidth]{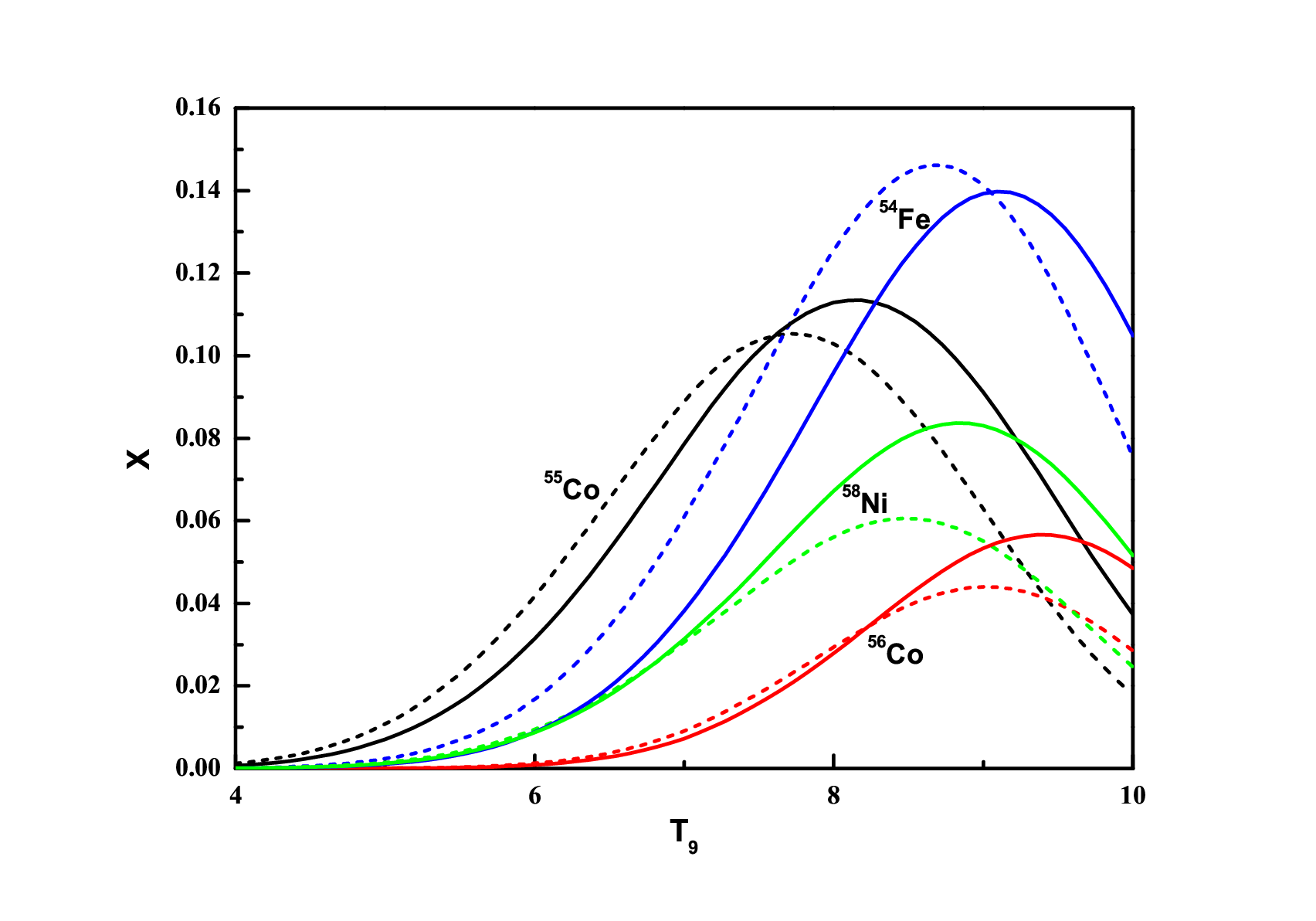}
	\vspace{-8mm}
	\caption{Same as Fig.~\ref{Ni56}, but for $^{54}$Fe, $^{55-56}$Co and $^{58}$Ni. }
	\label{MFvT}
\end{figure}

\begin{figure}[h]
	\centering
	\includegraphics[width=1.\textwidth]{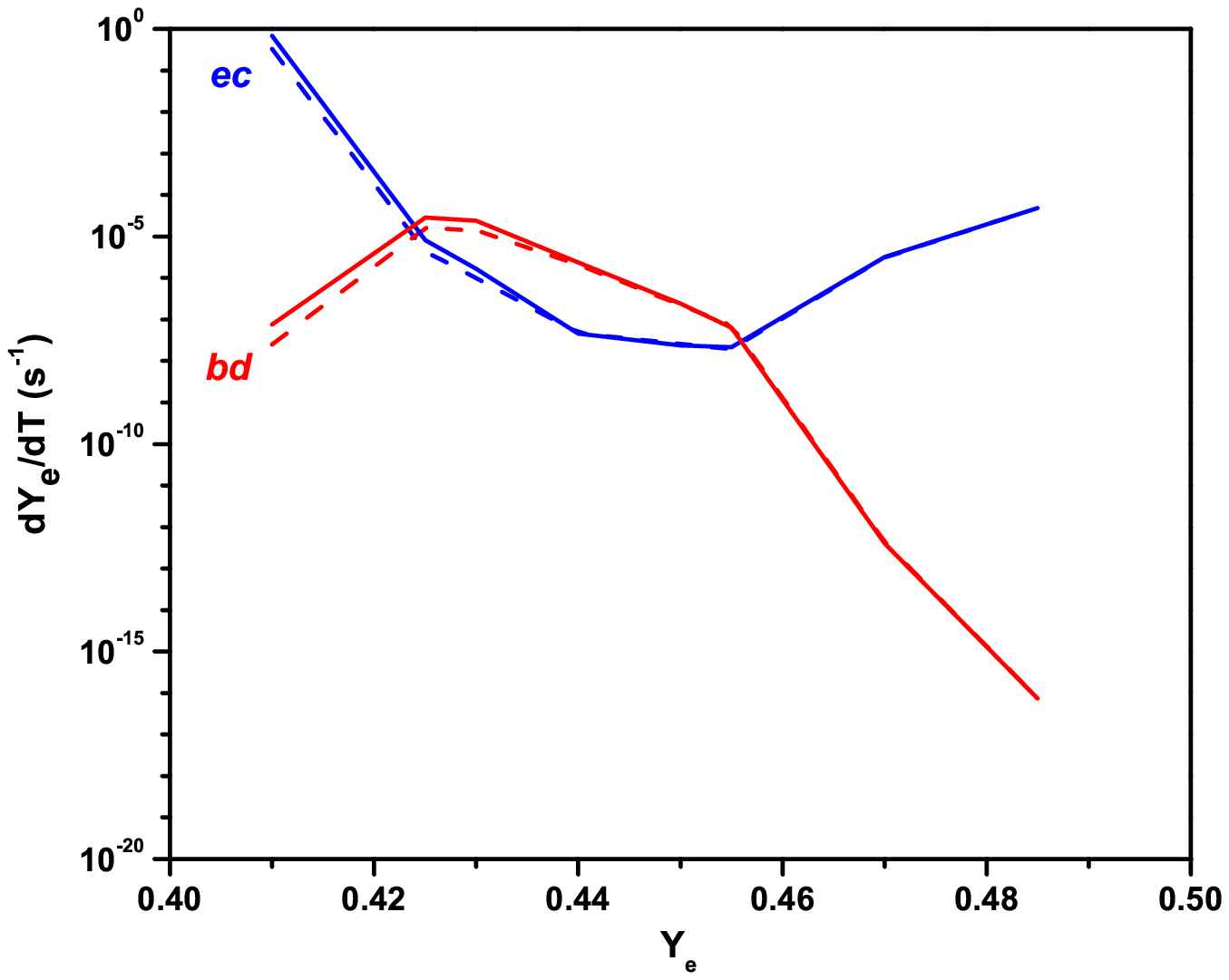}
	\vspace{-8mm}
	\caption{The total $\dot{Y}_e$ for \textit{ec} and \textit{bd} rates as a function of $Y_e$. The solid curves are the results computed with the Coulomb corrections (current work) while the dashed curves are for those without Coulomb corrections (Nab21).}
	\label{ECvBD}
\end{figure}

Recently, Nabi and collaborators \cite{Nab21} conducted  study of the presupernova evolution employing the pn-QRPA computed rates. The authors calculated the stellar weak rates, in a totally microscopic fashion without using the Brink-Axel hypothesis, and mass abundances for a total 728 nuclei with mass number A = 1–-100 and compiled a list of 50 most important
\textit{bd} and \textit{ec} nuclei that had the greatest impact on $Y_e$.
In this study we re-examine the rate of change of $Y_e$ during the presupernova evolution by coupling the pn-QRPA computed rates \cite{Nab21} (referred to as Nab21 here onwards) with newly computed mass abundances. The mass abundances, for the same pool of nuclei as Nab21, were computed with the incorporation of Coulomb corrections to the chemical potential of nuclear species, following the method of Bravo and García-Senz \cite{Bra99}.  \\
Fig.~\ref{F1.mf} displays our calculated mass abundances with (solid lines) and without (dashed lines) the Coulomb corrections to the chemical potential, for few abundant nuclei as a function of neutron excess ($\eta$) at $T$ = 8$\times$10$^9$ K and $\rho$ = 2$\times$10$^9$ g/cm$^3$.
The parameters ($T_9$, $\rho$ and $\eta$) were taken the same as used in Fig. 3 of Bravo and García-Senz \cite{Bra99}. The two figures compare well. The slight variations are presumably due to differences in the network used and the input nuclear data. It is noted that the Coulomb corrections alter the mass abundances of nuclei, albeit slightly. The mass abundances for Co isotopes are enhanced with the inclusion of Coulomb corrections in the whole range of $\eta$ = 0 - 0.08 ($Y_e$ = 0.50 - 0.46). For $^{55}$Co, the increment was up to $\sim$ 25 \% at $\eta$ = 0 ($Y_e$ = 0.50) and up to $\sim$ 40 \% at $\eta$ = 0.08. The calculated mass abundances are of the order of 10$^{-5}$ at $\eta$ = 0.08. This is the reason that the difference with  corrections incorporated cannot be seen clearly in the figure. The mass abundance for $^{55}$Fe was decreased at lower $\eta$ values while there was an increase at higher $\eta$ values once corrections were incorporated. The proton mass abundance was found rather insensitive to the corrections.  \\
Fig.~\ref{Ni56} shows the computed mass abundance for doubly magic nuclei $^{56}$Ni while Fig.~\ref{MFvT} depicts the results for $^{54}$Fe, $^{55-56}$Co and $^{58}$Ni as a function of $T_9$ (core temperature in units of 10$^{9}$ K) at fixed $\rho$ = 5$\times$10$^9$ g/cm$^3$ and $Y_e$ = 0.5. Since $\Gamma_i$ $\propto$ $Z_i^{5/3}$, the inclusion of Coulomb factor favors nuclei with large Z number, e.g. $^{56}$Ni \& $^{55}$Co (Fig.~\ref{Ni56} \& \ref{MFvT}), largely at the expense of some Fe nuclei and $^4$He. The computed abundance of $^{56}$Ni (Fig.~\ref{Ni56}) decreases with increase in temperature and results in increase of abundance of other nuclei ($^{55-56}$Co and $^{58}$Ni, Fig.~\ref{MFvT}). It can be seen that the calculated abundances are enhanced at large $T_9$ values and suppressed at lower values of temperature, once the Coulomb corrections are incorporated. This behaviour is in accordance with the effect depicted in Fig.~ 3, 4 \& 5 of Seitenzahl et al. \cite{Sei09}. One may expect noticeable increase in the total rate of change of $Y_e$ in NSE using the Coulomb corrections which we discuss later.\\
Tables (\ref{T1}-\ref{T8}) present our computed mass abundances (with Coulomb corrections) and the corresponding time rate of change of $Y_e$ ($\dot{Y}_e$), separately for \textit{bd} (Tables (\ref{T1}-\ref{T4})) and \textit{ec} (Tables (\ref{T5}-\ref{T8})). In each table, 30 nuclei have been sorted on the basis of their large contribution to $\dot{Y}_e$. The last column in these tables lists the order in which these nuclei were ranked by Nab21 on the basis of $\dot{Y}_e$. The stellar weak rates in Tables (\ref{T1}-\ref{T8}) were previously computed by Nabi et al. \cite{Nab21}, employing only allowed transitions. In comparison with Nab21 results, our computed mass abundances and the corresponding $\dot{Y}_e$ values are smaller for lower Z nuclei (Ca, Sc, Ti, V, Cr, etc.) and bigger for heavier nuclei (Ga, Ge, As, Se, Br, Kr, etc.). For example, our calculations are smaller by a factor of 2 for $^{53}$V, $^{56-57}$Cr and $^{50}$Sc and a factor 3 for $^{49,51}$Sc, $^{49}$Ca and $^{53}$Ti (see Tables (\ref{T1}-\ref{T8})). Similarly, our computed mass abundances are bigger than Nab21 results by a factor 2 for $^{75}$Ga, $^{64}$Ni, $^{69}$Zn and $^{59}$Fe and a factor 3 for $^{71-72}$Ga, $^{75}$Ge and $^{66}$Cu. The reduction,  up to a factor 4, was noted for low Z nuclei. Few exceptions for Ga and Ge isotopes were also noted where, despite high Z, the results were reduced up to an order of magnitude. On the other hand, our computed mass abundances were significantly enhanced for heavy nuclei at high density once the Coulomb corrections were incorporated. The enhancement factor was an order of magnitude for $^{86}$Br, $^{89}$Kr, $^{81}$Se, $^{60}$Co and $^{65}$Cu and up to 2 orders of magnitude for $^{59}$Co, $^{61}$Ni, $^{63}$Cu, $^{56}$Fe and $^{53}$Mn (see Tables (\ref{T1}, \ref{T5} \& \ref{T6})). This behaviour was expected since $\Gamma_i$ $\propto$ $Z_i^{5/3}$ and according to a previous study the corrections are relevant at higher density \cite{Bra99}. \\
Fig.~\ref{ECvBD} depicts the evolution of the temporal derivative of lepton-to-baryon fraction ($\dot{Y}_e$), with (solid curve) and without (dashed curve) the Coulomb corrections. The cup and cap-shaped curve represents the results for \textit{ec} and \textit{bd}, respectively. It is seen that the \textit{ec} rates predominate at the end values of $Y_e$, mainly because of the major contribution coming from $^{67}$Cu, $^{82}$Se, $^{59}$Co and $^{52}$V. The \textit{bd} rates dominate somewhere at the middle ($Y_e$ $\approx$ 0.440) because of the large contribution of $^{63}$Co, $^{49}$Sc, $^{67}$Ni and $^{57}$Fe to the total $\dot{Y}^{bd}_e$. Of special significance for the collapse simulators is the range of $Y_e$ where \textit{bd} rates surpass the corresponding \textit{ec} rates.  The total $\dot{Y}_e$ for \textit{bd} dominates the \textit{ec} results for the interval $Y_e$ $\approx$ 0.423-0.455, 3.2\% greater than the one previously proposed by Nab21 ($Y_e$ $\approx$ 0.424-0.455), once Coulomb corrections are incorporated. Because of the inverse dependence of $\Gamma_i$ on temperature, i-e $\Gamma_i$ $\propto$ $Z_i^{5/3}$/T, the corrections are  applied preferably at low core temperatures and high
densities. This fact is evident from Fig.~\ref{ECvBD} and the total $\dot{Y}_e$ values are high for higher values of $\rho$ (lower $Y_e$) with Coulomb corrections. At $Y_e$ = 0.425 ($\rho$ = 2.20$\times$10$^9$ g/cm$^3$), the magnitudes of $\dot{Y}_e$ for \textit{ec} and \textit{bd} are increased by 91\% and 80\%, respectively, after the incorporation of Coulomb corrections. The larger abundance of Co, Ni, Mn and some other isotopes could be the cause of this large increment (see Fig.~\ref{F1.mf} and Tables (\ref{T1}, \ref{T5} \& \ref{T6})).  \\
Finally, the contribution of each nucleus to the $\dot{Y}_{e}$ was averaged over the whole selected stellar trajectory to identify the most important weak interaction nuclei, using a ranking parameter ($\mathring{R}_p$) previously defined by Nab21 \cite{Nab21} as: 
\begin{equation}
	\mathring{R}_{p} = \left(\frac{\dot{Y}^{ec(bd)}_{e(i)}}{\sum\dot{Y}^{ec(bd)}_{e(i)}}\right)_{0.500 > {Y_e} > 0.400}	
\end{equation}
The nuclei that contribute the most to $\dot{Y_e}$ will have the highest $\mathring{R}_p$ value. We tabulate top 50 \textit{ec} and \textit{bd} nuclei based on $\mathring{R}_p$ in Table~\ref{T9}. Due to the enhanced abundances of nuclei at high density as a result of introducing Coulomb corrections, some new heavier nuclei (marked with an asterisk) make to the top 50 list. A total of 17 nuclei (11 \textit{ec} and 6 \textit{bd}) are debutants, not to be found in the list provided by Nab21.

\section{Summary and Conclusions}

In this study, we re-examined the rate of change of   ${Y}_e$ for  post silicon burning conditions during the presupernova evolution. We used the stellar weak rates computed by Nab21 employing the fully microscopic pn-QRPA model. The main reason for using the pn-QRPA stellar rates is that it adds to the reliability of our calculations since pn-QRPA does not use Brink's hypothesis to compute the GT strength distributions of parent excited states, as is the case with conventional stellar rate calculations. The mass abundances were computed, from Saha's equation assuming NSE, with the incorporation of Coulomb correction factor to the chemical potential, which according to a previous study are important at temperatures $T$ $\leq$ 5--10 GK and densities $\rho$ $\ge$ $10^8$ g/cm$^{3}$ \cite{Bra99}. \\
Our computed mass abundances and corresponding $\dot{Y}_e$ values were reduced only up to a few factor for low Z nuclei and enhanced up to orders of magnitude for heavy nuclei ($\Gamma_i$ $\propto$ $Z_i^{5/3}$), once Coulomb corrections were incorporated. Due to this significant enhancement, the total $\dot{Y}_e$ values for \textit{ec} and \textit{bd} were increased by 91\% and 80\%, respectively at $Y_e$ = 0.425 ($\rho$ = 2.20$\times$10$^9$ g/cm$^3$). Moreover, 11 \textit{ec} and 6 \textit{bd} nuclei debuted in our compiled list of top 50 important weak interaction nuclei (marked with an asterisk in Table \ref{T9}) due to enhanced abundances at high $\rho$ values. The interval where \textit{bd} rates surpass the competing \textit{ec} rates, was increased by 3.2 \% once Coulomb corrections were incorporated. \\
For future study, our selected ensemble of nuclei may be extended to include $Z >$ 40 nuclei since Coulomb corrections favour heavy Z number. We further plan to incorporate the conventional Debye length for computing the Coulomb corrections. Moreover, the role of forbidden transitions may be incorporated specially for neutron-rich cores.  \\ 
Unlike the massive stars (mass $>$ 8M$_\odot$) which goes through all burning phases (starting from hydrogen down to silicon burning) and synthesize nuclei up to the Fe core, only lighter nuclei can be synthesized in the Sun because of the previaling low temperatures. Much earlier Bahcall predicted neutrino fluxes and performed $\beta$-decay calculations for a number of nuclei \cite{Bah64}. Later a study of standard solar models and the uncertainties in predicted capture rates of solar neutrinos was performed \cite{Bah82}. However, the solar system may have more screening effects than that proposed by the  BBN theory because of the prevailing higher density. We plan to extend our study to the solar system and compare our calculations with the standard solar system in the near future.
 


\newpage

\newpage

\clearpage
\begin{table}[]
	\centering\scriptsize\caption{The mass fractions (\textit{X}), $\beta$-decay rates ($\lambda^{bd}$) and time rate of change of $Y_e$ for most relevant nuclei sorted by $\mid$$\dot{Y}^{bd}_e$$\mid$. The \textit{X} and $\dot{Y}^{bd}_e$ are compared with the Nab21 data.  The units of $T_9$, $\lambda^{bd}$, $\rho$ and $\dot{Y}^{bd}_e$ are \textit{GK}, $s^{-1}$, $g/cm^3$ and $s^{-1}$, respectively.   } \label{T1}
	\renewcommand{\arraystretch}{1.45}
	\begin{tabular}{cc|cc|c|cc|c}
		\hline
		\multicolumn{2}{}{}&\multicolumn{5}{c}{$\rho$ = 4.01(+10),~~~~~~~~~~~~$T_9$ = 7.33,~~~~~~~~~~~~$Y_e$ = 0.41}&\multicolumn{1}{}{} \\
		\hline
		\multicolumn{2}{}{}        & \multicolumn{2}{c}{Mass Fraction} & \multicolumn{1}{c}{$\lambda^{bd}$}      & \multicolumn{2}{c}{$\dot{Y}^{bd}_e$}  &            \\
		A  & Symbol & This work   & Nab21   & Nab21 & This work & Nab21 & Rank (Nab21) \\
		\hline
		85 & Se     & 1.14(-02)       & 6.70(-03)       & 3.16(-06)  & 4.24(-10)     & 2.49(-10)     & 19         \\
		87 & Se     & 5.01(-05)       & 6.12(-05)       & 5.82(-04)  & 3.35(-10)     & 4.09(-10)     & 10         \\
		84 & As     & 9.10(-04)       & 1.56(-03)       & 2.76(-05)  & 2.99(-10)     & 5.13(-10)     & 8          \\
		79 & Ga     & 2.19(-03)       & 9.81(-03)       & 9.40(-06)  & 2.61(-10)     & 1.17(-09)     & 4          \\
		67 & Co     & 6.26(-05)       & 6.75(-04)       & 2.09(-04)  & 1.96(-10)     & 2.11(-09)     & 2          \\
		49 & Ca     & 7.28(-04)       & 3.77(-02)       & 1.23(-05)  & 1.82(-10)     & 9.44(-09)     & 1          \\
		77 & Ga     & 1.37(-02)       & 2.95(-02)       & 9.16(-07)  & 1.63(-10)     & 3.51(-10)     & 12         \\
		78 & Ga     & 2.76(-03)       & 8.56(-03)       & 1.85(-06)  & 6.52(-11)     & 2.02(-10)     & 23         \\
		66 & Co     & 1.67(-04)       & 1.25(-03)       & 2.08(-05)  & 5.27(-11)     & 3.94(-10)     & 11         \\
		83 & As     & 2.67(-02)       & 3.16(-02)       & 1.54(-07)  & 4.94(-11)     & 5.86(-11)     & 43         \\
		68 & Co     & 4.87(-06)       & 7.57(-05)       & 6.41(-04)  & 4.59(-11)     & 7.14(-10)     & 7          \\
		86 & Br     & 2.02(-02)       & 3.95(-03)       & 1.95(-07)  & 4.56(-11)     & 8.94(-12)     & 76         \\
		86 & Se     & 1.56(-03)       & 1.32(-03)       & 2.32(-06)  & 4.20(-11)     & 3.56(-11)     & 48         \\
		80 & Ga     & 8.13(-05)       & 5.26(-04)       & 3.79(-05)  & 3.85(-11)     & 2.49(-10)     & 18         \\
		84 & Se     & 1.37(-01)       & 5.56(-02)       & 2.29(-08)  & 3.72(-11)     & 1.51(-11)     & 63         \\
		81 & Ga     & 3.27(-05)       & 3.05(-04)       & 8.75(-05)  & 3.53(-11)     & 3.29(-10)     & 14         \\
		50 & Sc     & 3.30(-04)       & 8.34(-03)       & 4.92(-06)  & 3.24(-11)     & 8.21(-10)     & 6          \\
		89 & Kr     & 4.07(-03)       & 5.43(-04)       & 6.30(-07)  & 2.88(-11)     & 3.84(-12)     & 88         \\
		50 & Ca     & 7.31(-05)       & 5.46(-03)       & 1.90(-05)  & 2.78(-11)     & 2.08(-09)     & 3          \\
		81 & Ge     & 9.05(-03)       & 2.11(-02)       & 2.38(-07)  & 2.66(-11)     & 6.21(-11)     & 42         \\
		65 & Co     & 7.67(-04)       & 3.96(-03)       & 2.13(-06)  & 2.51(-11)     & 1.30(-10)     & 29         \\
		62 & Fe     & 1.98(-03)       & 1.19(-02)       & 7.10(-07)  & 2.27(-11)     & 1.36(-10)     & 27         \\
		73 & Cu     & 1.98(-04)       & 1.45(-03)       & 7.52(-06)  & 2.04(-11)     & 1.49(-10)     & 26         \\
		82 & Ge     & 3.53(-03)       & 1.19(-02)       & 4.65(-07)  & 2.00(-11)     & 6.74(-11)     & 39         \\
		58 & Cr     & 2.20(-04)       & 3.44(-03)       & 5.04(-06)  & 1.91(-11)     & 2.99(-10)     & 15         \\
		59 & Cr     & 9.49(-06)       & 2.14(-04)       & 1.16(-04)  & 1.87(-11)     & 4.21(-10)     & 9          \\
		72 & Cu     & 4.03(-04)       & 2.05(-03)       & 3.27(-06)  & 1.83(-11)     & 9.30(-11)     & 33         \\
		87 & Br     & 3.07(-03)       & 8.70(-04)       & 5.14(-07)  & 1.82(-11)     & 5.14(-12)     & 83         \\
		88 & Br     & 2.64(-04)       & 1.08(-04)       & 5.36(-06)  & 1.61(-11)     & 6.58(-12)     & 79         \\
		82 & As     & 2.15(-02)       & 1.76(-02)       & 6.11(-08)  & 1.60(-11)     & 1.31(-11)     & 65        \\
		\hline
	\end{tabular}
\end{table}
\begin{table}[]
	\centering\scriptsize\caption{Same as Table \ref{T1} but for physical conditions listed below. } \label{T2}
	\renewcommand{\arraystretch}{1.45}
	\begin{tabular}{cc|cc|c|cc|c}
		\hline
		\multicolumn{2}{}{}&\multicolumn{5}{c}{$\rho$ = 1.06(+09),~~~~~~~~~~~~$T_9$ = 4.93,~~~~~~~~~~~~$Y_e$ = 0.43}&\multicolumn{1}{}{} \\
		\hline
		\multicolumn{2}{}{}        & \multicolumn{2}{c}{Mass Fraction} & \multicolumn{1}{c}{$\lambda^{bd}$}      & \multicolumn{2}{c}{$\dot{Y}^{bd}_e$}  &            \\
		A  & Symbol & This work   & Nab21   & Nab21 & This work & Nab21 & Rank (Nab21) \\
		\hline
		67 & Ni & 3.20(-02) & 3.13(-02) & 4.12(-03) & 1.97(-06) & 1.93(-06) & 1  \\
		65 & Co & 2.31(-04) & 2.90(-04) & 4.16(-01) & 1.48(-06) & 1.85(-06) & 2  \\
		64 & Co & 6.45(-04) & 7.14(-04) & 8.87(-02) & 8.94(-07) & 9.90(-07) & 4  \\
		59 & Mn & 6.35(-04) & 9.82(-04) & 6.08(-02) & 6.55(-07) & 1.01(-06) & 3  \\
		63 & Co & 5.27(-03) & 5.15(-03) & 7.23(-03) & 6.04(-07) & 5.91(-07) & 8  \\
		62 & Fe & 6.53(-03) & 9.15(-03) & 3.03(-03) & 3.20(-07) & 4.48(-07) & 13 \\
		61 & Fe & 3.47(-03) & 4.30(-03) & 5.47(-03) & 3.12(-07) & 3.86(-07) & 14 \\
		57 & Cr & 3.35(-04) & 6.38(-04) & 4.89(-02) & 2.87(-07) & 5.47(-07) & 9  \\
		49 & Sc & 2.58(-03) & 6.65(-03) & 4.92(-03) & 2.59(-07) & 6.68(-07) & 7  \\
		51 & Ti & 1.12(-02) & 2.43(-02) & 1.10(-03) & 2.40(-07) & 5.22(-07) & 10 \\
		63 & Fe & 3.54(-05) & 5.62(-05) & 4.24(-01) & 2.38(-07) & 3.78(-07) & 15 \\
		50 & Sc & 3.04(-04) & 8.86(-04) & 3.85(-02) & 2.34(-07) & 6.81(-07) & 6  \\
		51 & Sc & 2.27(-05) & 7.49(-05) & 5.18(-01) & 2.30(-07) & 7.60(-07) & 5  \\
		58 & Cr & 1.46(-04) & 3.15(-04) & 8.28(-02) & 2.09(-07) & 4.50(-07) & 12 \\
		69 & Cu & 3.12(-03) & 2.36(-03) & 4.46(-03) & 2.02(-07) & 1.52(-07) & 21 \\
		75 & Ga & 1.20(-03) & 6.65(-04) & 1.10(-02) & 1.76(-07) & 9.75(-08) & 24 \\
		71 & Cu & 5.78(-05) & 5.60(-05) & 2.12(-01) & 1.73(-07) & 1.67(-07) & 20 \\
		53 & Ti & 1.75(-04) & 4.88(-04) & 5.05(-02) & 1.67(-07) & 4.65(-07) & 11 \\
		66 & Co & 3.39(-06) & 4.82(-06) & 2.86(-00) & 1.47(-07) & 2.09(-07) & 17 \\
		77 & Ga & 8.33(-05) & 5.93(-05) & 1.25(-01) & 1.35(-07) & 9.64(-08) & 25 \\
		60 & Mn & 3.96(-05) & 6.93(-05) & 1.69(-01) & 1.12(-07) & 1.96(-07) & 18 \\
		70 & Cu & 9.17(-05) & 7.85(-05) & 7.62(-02) & 9.99(-08) & 8.54(-08) & 26 \\
		54 & V  & 1.85(-04) & 3.79(-04) & 2.77(-02) & 9.51(-08) & 1.95(-07) & 19 \\
		55 & Cr & 5.62(-03) & 8.35(-03) & 9.12(-04) & 9.32(-08) & 1.39(-07) & 22 \\
		49 & Ca & 3.34(-04) & 1.29(-03) & 1.17(-02) & 8.00(-08) & 3.09(-07) & 16 \\
		56 & Cr & 2.14(-02) & 3.60(-02) & 1.23(-04) & 4.69(-08) & 7.89(-08) & 27 \\
		66 & Ni & 3.33(-01) & 2.88(-01) & 9.23(-06) & 4.66(-08) & 4.03(-08) & 33 \\
		68 & Ni & 1.89(-02) & 2.09(-02) & 1.58(-04) & 4.40(-08) & 4.88(-08) & 29 \\
		55 & V  & 8.57(-05) & 1.99(-04) & 2.72(-02) & 4.23(-08) & 9.82(-08) & 23 \\
		58 & Mn & 8.63(-04) & 1.18(-03) & 2.31(-03) & 3.43(-08) & 4.69(-08) & 30\\
		\hline
	\end{tabular}
\end{table}
\begin{table}[]
	\centering\scriptsize\caption{Same as Table \ref{T1} but for physical conditions listed below.  } \label{T3}
	\renewcommand{\arraystretch}{1.45}
	\begin{tabular}{cc|cc|c|cc|c}
		\hline
		\multicolumn{2}{}{}&\multicolumn{5}{c}{$\rho$ = 1.45(+08),~~~~~~~~~~~~$T_9$ = 3.8,~~~~~~~~~~~~$Y_e$ = 0.45}&\multicolumn{1}{}{} \\
		\hline
		\multicolumn{2}{}{}        & \multicolumn{2}{c}{Mass Fraction} & \multicolumn{1}{c}{$\lambda^{bd}$}      & \multicolumn{2}{c}{$\dot{Y}^{bd}_e$}  &            \\
		A  & Symbol & This work   & Nab21   & Nab21 & This work & Nab21 & Rank (Nab21) \\
		\hline
		57 & Mn & 2.13(-04) & 2.27(-04) & 4.92(-03) & 1.84(-08) & 1.96(-08) & 1  \\
		55 & Cr & 1.76(-04) & 2.08(-04) & 4.10(-03) & 1.31(-08) & 1.55(-08) & 2  \\
		49 & Sc & 3.69(-06) & 5.71(-06) & 6.55(-02) & 4.93(-09) & 7.63(-09) & 3  \\
		52 & V  & 5.25(-05) & 7.18(-05) & 3.95(-03) & 3.99(-09) & 5.46(-09) & 4  \\
		61 & Co & 1.18(-03) & 9.91(-04) & 1.64(-04) & 3.19(-09) & 2.67(-09) & 6  \\
		51 & Ti & 5.83(-05) & 8.32(-05) & 2.28(-03) & 2.60(-09) & 3.71(-09) & 5  \\
		63 & Co & 2.09(-06) & 1.60(-06) & 4.86(-02) & 1.62(-09) & 1.23(-09) & 9  \\
		53 & V  & 1.80(-05) & 2.35(-05) & 4.55(-03) & 1.54(-09) & 2.01(-09) & 7  \\
		54 & Cr & 1.36(-01) & 1.69(-01) & 5.05(-07) & 1.27(-09) & 1.58(-09) & 8  \\
		59 & Fe & 2.06(-03) & 1.96(-03) & 3.52(-05) & 1.23(-09) & 1.17(-09) & 10 \\
		58 & Fe & 4.41(-01) & 4.38(-01) & 7.91(-08) & 6.02(-10) & 5.97(-10) & 11 \\
		62 & Co & 5.35(-06) & 4.28(-06) & 4.42(-03) & 3.81(-10) & 3.05(-10) & 12 \\
		65 & Ni & 4.48(-05) & 3.00(-05) & 4.45(-04) & 3.07(-10) & 2.05(-10) & 14 \\
		58 & Mn & 7.74(-07) & 7.87(-07) & 1.97(-02) & 2.63(-10) & 2.67(-10) & 13 \\
		61 & Fe & 4.46(-07) & 3.86(-07) & 2.20(-02) & 1.61(-10) & 1.39(-10) & 15 \\
		59 & Mn & 2.10(-08) & 2.04(-08) & 1.95(-01) & 6.94(-11) & 6.74(-11) & 18 \\
		56 & Cr & 4.01(-05) & 4.53(-05) & 9.06(-05) & 6.48(-11) & 7.33(-11) & 17 \\
		60 & Fe & 2.08(-03) & 1.88(-03) & 1.81(-06) & 6.26(-11) & 5.66(-11) & 19 \\
		50 & Sc & 4.00(-09) & 5.91(-09) & 6.70(-01) & 5.35(-11) & 7.92(-11) & 16 \\
		56 & Mn & 3.37(-04) & 3.75(-04) & 8.41(-06) & 5.06(-11) & 5.63(-11) & 20 \\
		64 & Ni & 2.85(-02) & 1.99(-02) & 7.28(-08) & 3.24(-11) & 2.27(-11) & 22 \\
		67 & Ni & 3.81(-08) & 2.32(-08) & 4.74(-02) & 2.70(-11) & 1.64(-11) & 24 \\
		55 & Mn & 9.57(-03) & 1.12(-02) & 1.46(-07) & 2.53(-11) & 2.95(-11) & 21 \\
		54 & V  & 1.56(-08) & 1.94(-08) & 6.03(-02) & 1.74(-11) & 2.17(-11) & 23 \\
		64 & Co & 2.28(-09) & 1.67(-09) & 3.57(-01) & 1.27(-11) & 9.30(-12) & 26 \\
		67 & Cu & 1.86(-05) & 1.08(-05) & 3.58(-05) & 9.93(-12) & 5.77(-12) & 30 \\
		63 & Ni & 2.40(-03) & 1.76(-03) & 2.43(-07) & 9.23(-12) & 6.76(-12) & 28 \\
		53 & Cr & 8.52(-03) & 1.11(-02) & 5.62(-08) & 9.04(-12) & 1.17(-11) & 25 \\
		57 & Cr & 3.64(-09) & 3.93(-09) & 1.24(-01) & 7.90(-12) & 8.54(-12) & 27 \\
		60 & Co & 2.29(-04) & 2.01(-04) & 1.80(-06) & 6.89(-12) & 6.04(-12) & 29 \\
		\hline
	\end{tabular}
\end{table}
\begin{table}[]
	\centering\scriptsize\caption{Same as Table \ref{T1} but for physical conditions listed below. } \label{T4}
	\renewcommand{\arraystretch}{1.45}
	\begin{tabular}{cc|cc|c|cc|c}
		\hline
		\multicolumn{2}{}{}&\multicolumn{5}{c}{$\rho$ = 5.86(+07),~~~~~~~~~~~~$T_9$ = 3.4,~~~~~~~~~~~~$Y_e$ = 0.47}&\multicolumn{1}{}{} \\
		\hline
		\multicolumn{2}{}{}        & \multicolumn{2}{c}{Mass Fraction} & \multicolumn{1}{c}{$\lambda^{bd}$}      & \multicolumn{2}{c}{$\dot{Y}^{bd}_e$}  &            \\
		A  & Symbol & This work   & Nab21   & Nab21 & This work & Nab21 & Rank (Nab21) \\
		\hline
		55 & Mn & 8.25(-05) & 9.08(-05) & 1.08(-07) & 1.62(-13) & 1.78(-13) & 1  \\
		57 & Fe & 1.62(-04) & 1.61(-04) & 3.34(-08) & 9.49(-14) & 9.45(-14) & 2  \\
		58 & Fe & 8.50(-05) & 8.28(-05) & 2.90(-08) & 4.26(-14) & 4.14(-14) & 3  \\
		52 & V  & 1.82(-10) & 2.35(-10) & 8.71(-03) & 3.05(-14) & 3.93(-14) & 4  \\
		53 & Cr & 2.97(-05) & 3.60(-05) & 3.75(-08) & 2.10(-14) & 2.55(-14) & 5  \\
		54 & Cr & 3.84(-06) & 4.54(-06) & 2.02(-07) & 1.44(-14) & 1.70(-14) & 6  \\
		59 & Co & 1.52(-04) & 1.36(-04) & 3.86(-09) & 9.95(-15) & 8.90(-15) & 7  \\
		61 & Co & 1.25(-09) & 1.07(-09) & 2.86(-04) & 5.87(-15) & 5.01(-15) & 8  \\
		57 & Mn & 3.09(-11) & 3.26(-11) & 7.43(-03) & 4.03(-15) & 4.24(-15) & 9  \\
		56 & Fe & 5.19(-01) & 5.29(-01) & 4.32(-13) & 4.00(-15) & 4.08(-15) & 10 \\
		54 & Mn & 1.29(-04) & 1.45(-04) & 1.12(-09) & 2.67(-15) & 3.01(-15) & 11 \\
		53 & Mn & 5.54(-03) & 6.39(-03) & 1.76(-11) & 1.84(-15) & 2.12(-15) & 12 \\
		60 & Co & 3.48(-08) & 3.04(-08) & 2.03(-06) & 1.18(-15) & 1.03(-15) & 15 \\
		56 & Mn & 8.70(-09) & 9.37(-09) & 7.41(-06) & 1.15(-15) & 1.24(-15) & 13 \\
		55 & Cr & 1.02(-11) & 1.18(-11) & 4.84(-03) & 9.02(-16) & 1.04(-15) & 14 \\
		52 & Cr & 2.60(-02) & 3.22(-02) & 1.48(-12) & 7.42(-16) & 9.19(-16) & 16 \\
		59 & Fe & 9.66(-10) & 9.19(-10) & 4.28(-05) & 7.00(-16) & 6.66(-16) & 17 \\
		57 & Co & 9.62(-03) & 9.01(-03) & 2.27(-12) & 3.83(-16) & 3.59(-16) & 18 \\
		55 & Fe & 2.75(-02) & 2.87(-02) & 5.01(-13) & 2.51(-16) & 2.61(-16) & 20 \\
		51 & V  & 1.86(-06) & 2.46(-06) & 5.96(-09) & 2.18(-16) & 2.87(-16) & 19 \\
		51 & Cr & 7.65(-05) & 9.69(-05) & 1.05(-10) & 1.57(-16) & 1.99(-16) & 21 \\
		61 & Ni & 6.55(-05) & 5.22(-05) & 5.09(-11) & 5.47(-17) & 4.36(-17) & 24 \\
		63 & Ni & 6.50(-09) & 4.94(-09) & 4.50(-07) & 4.64(-17) & 3.53(-17) & 26 \\
		53 & V  & 3.26(-13) & 4.11(-13) & 6.95(-03) & 4.28(-17) & 5.39(-17) & 22 \\
		58 & Co & 6.90(-05) & 6.31(-05) & 2.96(-11) & 3.52(-17) & 3.22(-17) & 27 \\
		49 & Ti & 8.65(-09) & 1.23(-08) & 1.96(-07) & 3.47(-17) & 4.94(-17) & 23 \\
		51 & Ti & 4.77(-13) & 6.49(-13) & 2.89(-03) & 2.70(-17) & 3.68(-17) & 25 \\
		62 & Ni & 2.24(-04) & 1.75(-04) & 7.36(-12) & 2.66(-17) & 2.07(-17) & 30 \\
		50 & V  & 1.12(-07) & 1.51(-07) & 7.87(-09) & 1.76(-17) & 2.38(-17) & 28 \\
		49 & Sc & 9.33(-15) & 1.36(-14) & 8.30(-02) & 1.58(-17) & 2.31(-17) & 29 \\
		\hline
	\end{tabular}
\end{table}
\begin{table}[]
	\centering\scriptsize\caption{Same as Table \ref{T1} but for \textit{ec} and physical conditions listed below.  } \label{T5}
	\renewcommand{\arraystretch}{1.45}
	\begin{tabular}{cc|cc|c|cc|c}
		\hline
		\multicolumn{2}{}{}&\multicolumn{5}{c}{$\rho$ = 2.20(+09),~~~~~~~~~~~~$T_9$ = 5.39,~~~~~~~~~~~~$Y_e$ = 0.425}&\multicolumn{1}{}{} \\
		\hline
		\multicolumn{2}{}{}        & \multicolumn{2}{c}{Mass Fraction} & \multicolumn{1}{c}{$\lambda^{ec}$}      & \multicolumn{2}{c}{$\dot{Y}^{ec}_e$}  &            \\
		A  & Symbol & This work   & Nab21   & Nab21 & This work & Nab21 & Rank (Nab21) \\
		\hline
		56 & Mn & 3.11(-04) & 2.85(-04) & 7.45(-01) & 4.13(-04) & 3.79(-04) & 1  \\
		67 & Cu & 7.75(-03) & 4.02(-03) & 1.33(-01) & 1.54(-05) & 7.98(-06) & 3  \\
		60 & Co & 5.01(-05) & 2.58(-05) & 1.19(-01) & 9.92(-06) & 5.11(-06) & 5  \\
		52 & V  & 3.89(-04) & 6.01(-04) & 1.32(-00) & 9.91(-06) & 1.53(-05) & 2  \\
		49 & Sc & 3.07(-03) & 9.28(-03) & 3.68(-02) & 2.30(-06) & 6.97(-06) & 4  \\
		66 & Cu & 4.47(-04) & 1.87(-04) & 3.04(-01) & 2.06(-06) & 8.61(-07) & 9  \\
		68 & Cu & 3.18(-03) & 2.05(-03) & 3.27(-02) & 1.53(-06) & 9.83(-07) & 8  \\
		65 & Cu & 1.39(-04) & 4.69(-05) & 6.40(-01) & 1.37(-06) & 4.61(-07) & 18 \\
		73 & Ga & 1.97(-03) & 7.87(-04) & 4.62(-02) & 1.25(-06) & 4.98(-07) & 16 \\
		64 & Ni & 1.02(-01) & 5.90(-02) & 6.76(-04) & 1.08(-06) & 6.23(-07) & 13 \\
		72 & Ga & 2.91(-04) & 9.36(-05) & 2.61(-01) & 1.06(-06) & 3.39(-07) & 21 \\
		75 & Ge & 1.87(-03) & 5.17(-04) & 3.89(-02) & 9.72(-07) & 2.68(-07) & 23 \\
		54 & Cr & 2.07(-02) & 2.48(-02) & 1.82(-03) & 6.97(-07) & 8.37(-07) & 10 \\
		51 & Ti & 1.09(-02) & 2.64(-02) & 2.96(-03) & 6.33(-07) & 1.54(-06) & 6  \\
		65 & Ni & 2.88(-02) & 2.07(-02) & 1.32(-03) & 5.84(-07) & 4.19(-07) & 20 \\
		64 & Co & 1.48(-03) & 1.80(-03) & 2.44(-02) & 5.64(-07) & 6.86(-07) & 11 \\
		69 & Cu & 6.52(-03) & 5.20(-03) & 5.94(-03) & 5.62(-07) & 4.48(-07) & 19 \\
		63 & Ni & 1.58(-03) & 7.37(-04) & 2.23(-02) & 5.58(-07) & 2.61(-07) & 25 \\
		71 & Ga & 1.49(-04) & 3.86(-05) & 2.47(-01) & 5.18(-07) & 1.34(-07) & 32 \\
		57 & Mn & 2.49(-03) & 2.84(-03) & 1.12(-02) & 4.92(-07) & 5.60(-07) & 14 \\
		61 & Co & 1.88(-03) & 1.20(-03) & 1.50(-02) & 4.61(-07) & 2.94(-07) & 22 \\
		55 & Cr & 4.68(-03) & 6.96(-03) & 4.24(-03) & 3.60(-07) & 5.36(-07) & 15 \\
		50 & Sc & 6.88(-04) & 2.58(-03) & 2.61(-02) & 3.58(-07) & 1.35(-06) & 7  \\
		69 & Zn & 2.58(-04) & 9.54(-05) & 8.49(-02) & 3.18(-07) & 1.17(-07) & 35 \\
		81 & Se & 3.43(-03) & 6.71(-04) & 6.67(-03) & 2.83(-07) & 5.52(-08) & 40 \\
		48 & Sc & 1.63(-05) & 3.97(-05) & 8.02(-01) & 2.72(-07) & 6.64(-07) & 12 \\
		59 & Fe & 5.79(-03) & 4.97(-03) & 2.64(-03) & 2.59(-07) & 2.23(-07) & 27 \\
		74 & Ga & 6.85(-04) & 3.39(-04) & 2.64(-02) & 2.45(-07) & 1.21(-07) & 34 \\
		53 & V  & 2.03(-03) & 3.89(-03) & 6.31(-03) & 2.42(-07) & 4.63(-07) & 17 \\
		53 & Cr & 2.53(-04) & 2.45(-04) & 4.68(-02) & 2.23(-07) & 2.16(-07) & 28\\
		\hline
	\end{tabular}
\end{table}
\begin{table}[]
	\centering\scriptsize\caption{Same as Table \ref{T1} but for \textit{ec} and physical conditions listed below.   } \label{T6}
	\renewcommand{\arraystretch}{1.45}
	\begin{tabular}{cc|cc|c|cc|c}
		\hline
		\multicolumn{2}{}{}&\multicolumn{5}{c}{$\rho$ = 3.30(+08),~~~~~~~~~~~~$T_9$ = 4.24,~~~~~~~~~~~~$Y_e$ = 0.44}&\multicolumn{1}{}{} \\
		\hline
		\multicolumn{2}{}{}        & \multicolumn{2}{c}{Mass Fraction} & \multicolumn{1}{c}{$\lambda^{ec}$}      & \multicolumn{2}{c}{$\dot{Y}^{ec}_e$}  &            \\
		A  & Symbol & This work   & Nab21   & Nab21 & This work & Nab21 & Rank (Nab21) \\
		\hline
		60 & Co & 1.52(-04) & 1.00(-05) & 1.60(-02) & 4.05(-08) & 2.67(-09) & 2  \\
		65 & Cu & 3.74(-04) & 2.93(-05) & 5.18(-03) & 2.97(-08) & 2.33(-09) & 3  \\
		67 & Cu & 2.53(-03) & 2.53(-03) & 3.26(-04) & 1.23(-08) & 1.23(-08) & 1  \\
		56 & Mn & 6.05(-04) & 1.16(-04) & 8.15(-04) & 8.81(-09) & 1.68(-09) & 4  \\
		53 & Cr & 2.05(-03) & 1.84(-04) & 1.23(-04) & 4.77(-09) & 4.27(-10) & 9  \\
		59 & Co & 1.94(-04) & 3.59(-06) & 1.19(-03) & 3.93(-09) & 7.27(-11) & 26 \\
		57 & Fe & 6.87(-04) & 2.17(-05) & 2.52(-04) & 3.03(-09) & 9.60(-11) & 24 \\
		63 & Ni & 5.33(-03) & 7.28(-04) & 3.40(-05) & 2.88(-09) & 3.94(-10) & 11 \\
		51 & V  & 8.00(-04) & 1.19(-04) & 1.82(-04) & 2.85(-09) & 4.23(-10) & 10 \\
		49 & Ti & 3.60(-05) & 8.75(-06) & 3.69(-03) & 2.71(-09) & 6.59(-10) & 7  \\
		52 & V  & 3.18(-04) & 1.69(-04) & 3.94(-04) & 2.41(-09) & 1.28(-09) & 5  \\
		55 & Mn & 1.37(-03) & 7.34(-05) & 9.51(-05) & 2.37(-09) & 1.27(-10) & 20 \\
		66 & Cu & 2.07(-04) & 5.79(-05) & 6.32(-04) & 1.98(-09) & 5.55(-10) & 8  \\
		61 & Ni & 2.96(-05) & 3.17(-07) & 4.00(-03) & 1.94(-09) & 2.08(-11) & 37 \\
		61 & Co & 4.58(-03) & 1.08(-03) & 1.63(-05) & 1.23(-09) & 2.89(-10) & 13 \\
		54 & Cr & 1.68(-01) & 5.38(-02) & 3.03(-07) & 9.43(-10) & 3.02(-10) & 12 \\
		63 & Cu & 9.84(-07) & 6.05(-09) & 4.17(-02) & 6.51(-10) & 4.01(-12) & 55 \\
		62 & Ni & 3.37(-02) & 1.29(-03) & 1.17(-06) & 6.37(-10) & 2.44(-11) & 36 \\
		67 & Zn & 2.80(-06) & 1.25(-07) & 1.43(-02) & 5.99(-10) & 2.67(-11) & 33 \\
		64 & Cu & 2.57(-06) & 5.64(-08) & 1.42(-02) & 5.70(-10) & 1.25(-11) & 38 \\
		56 & Fe & 6.76(-04) & 5.99(-06) & 4.16(-05) & 5.02(-10) & 4.45(-12) & 52 \\
		58 & Fe & 1.75(-01) & 1.98(-02) & 1.58(-07) & 4.79(-10) & 5.41(-11) & 31 \\
		64 & Ni & 3.29(-01) & 1.61(-01) & 7.50(-08) & 3.86(-10) & 1.88(-10) & 14 \\
		71 & Ga & 1.91(-05) & 6.11(-06) & 1.17(-03) & 3.15(-10) & 1.01(-10) & 23 \\
		54 & Mn & 4.69(-06) & 7.03(-08) & 3.24(-03) & 2.81(-10) & 4.21(-12) & 54 \\
		69 & Zn & 4.96(-05) & 2.81(-05) & 3.70(-04) & 2.66(-10) & 1.51(-10) & 15 \\
		53 & Mn & 1.49(-07) & 6.28(-10) & 9.29(-02) & 2.62(-10) & 1.10(-12) & 66 \\
		59 & Fe & 1.24(-02) & 5.00(-03) & 1.23(-06) & 2.59(-10) & 1.04(-10) & 22 \\
		57 & Mn & 2.71(-03) & 1.85(-03) & 4.63(-06) & 2.20(-10) & 1.50(-10) & 16 \\
		49 & Sc & 5.50(-04) & 2.77(-03) & 1.74(-05) & 1.95(-10) & 9.81(-10) & 6 \\
		\hline
	\end{tabular}
\end{table}
\begin{table}[]
	\centering\scriptsize\caption{Same as Table \ref{T1} but for \textit{ec} and physical conditions listed below.   } \label{T7}
	\renewcommand{\arraystretch}{1.45}
	\begin{tabular}{cc|cc|c|cc|c}
		\hline
		\multicolumn{2}{}{}&\multicolumn{5}{c}{$\rho$ = 1.07(+08),~~~~~~~~~~~~$T_9$ = 3.65,~~~~~~~~~~~~$Y_e$ = 0.455}&\multicolumn{1}{}{} \\
		\hline
		\multicolumn{2}{}{}        & \multicolumn{2}{c}{Mass Fraction} & \multicolumn{1}{c}{$\lambda^{ec}$}      & \multicolumn{2}{c}{$\dot{Y}^{ec}_e$}  &            \\
		A  & Symbol & This work   & Nab21   & Nab21 & This work & Nab21 & Rank (Nab21) \\
		\hline
		53 & Mn & 1.61(-04) & 1.82(-04) & 8.87(-03) & 2.69(-08) & 3.05(-08) & 1  \\
		55 & Fe & 3.70(-04) & 3.73(-04) & 2.70(-03) & 1.81(-08) & 1.83(-08) & 2  \\
		61 & Ni & 1.21(-03) & 9.43(-04) & 3.08(-04) & 6.09(-09) & 4.75(-09) & 4  \\
		59 & Co & 5.02(-03) & 4.48(-03) & 6.79(-05) & 5.78(-09) & 5.16(-09) & 3  \\
		56 & Fe & 3.02(-01) & 3.06(-01) & 7.69(-07) & 4.15(-09) & 4.20(-09) & 5  \\
		63 & Cu & 5.04(-05) & 3.41(-05) & 4.33(-03) & 3.46(-09) & 2.34(-09) & 8  \\
		57 & Co & 7.26(-05) & 6.44(-05) & 2.21(-03) & 2.82(-09) & 2.50(-09) & 7  \\
		51 & Cr & 5.53(-06) & 7.01(-06) & 2.01(-02) & 2.18(-09) & 2.77(-09) & 6  \\
		57 & Fe & 1.01(-02) & 1.02(-02) & 9.68(-06) & 1.71(-09) & 1.74(-09) & 9  \\
		54 & Mn & 2.95(-04) & 3.35(-04) & 1.87(-04) & 1.02(-09) & 1.16(-09) & 10 \\
		59 & Ni & 1.21(-05) & 9.37(-06) & 4.86(-03) & 9.97(-10) & 7.73(-10) & 11 \\
		58 & Co & 4.43(-05) & 3.94(-05) & 1.09(-03) & 8.34(-10) & 7.42(-10) & 12 \\
		60 & Co & 1.25(-04) & 1.12(-04) & 3.78(-04) & 7.86(-10) & 7.04(-10) & 13 \\
		55 & Mn & 1.03(-02) & 1.18(-02) & 3.23(-06) & 6.07(-10) & 6.93(-10) & 14 \\
		60 & Ni & 2.06(-02) & 1.60(-02) & 1.50(-06) & 5.16(-10) & 4.01(-10) & 16 \\
		53 & Cr & 7.59(-03) & 9.68(-03) & 3.15(-06) & 4.51(-10) & 5.75(-10) & 15 \\
		61 & Cu & 7.62(-08) & 5.12(-08) & 3.40(-01) & 4.24(-10) & 2.85(-10) & 18 \\
		65 & Cu & 7.64(-05) & 5.20(-05) & 2.55(-04) & 3.00(-10) & 2.04(-10) & 21 \\
		49 & V  & 1.49(-07) & 2.10(-07) & 8.41(-02) & 2.56(-10) & 3.60(-10) & 17 \\
		54 & Fe & 4.10(-05) & 4.12(-05) & 3.32(-04) & 2.52(-10) & 2.53(-10) & 19 \\
		52 & Cr & 6.76(-02) & 8.59(-02) & 1.32(-07) & 1.71(-10) & 2.17(-10) & 20 \\
		51 & V  & 1.13(-03) & 1.59(-03) & 6.21(-06) & 1.37(-10) & 1.94(-10) & 22 \\
		58 & Ni & 3.98(-06) & 3.07(-06) & 1.84(-03) & 1.26(-10) & 9.73(-11) & 25 \\
		48 & V  & 6.88(-11) & 9.62(-11) & 8.05(-01) & 1.15(-10) & 1.61(-10) & 23 \\
		50 & V  & 1.52(-06) & 2.14(-06) & 2.74(-03) & 8.33(-11) & 1.17(-10) & 24 \\
		64 & Cu & 4.80(-06) & 3.26(-06) & 7.96(-04) & 5.97(-11) & 4.05(-11) & 28 \\
		49 & Ti & 1.42(-05) & 2.21(-05) & 1.99(-04) & 5.77(-11) & 8.96(-11) & 26 \\
		56 & Co & 4.94(-08) & 4.37(-08) & 5.87(-02) & 5.18(-11) & 4.58(-11) & 27 \\
		64 & Zn & 8.58(-07) & 4.97(-07) & 2.54(-03) & 3.40(-11) & 1.97(-11) & 31 \\
		62 & Ni & 2.03(-01) & 1.59(-01) & 9.89(-09) & 3.24(-11) & 2.54(-11) & 29\\
		\hline
	\end{tabular}
\end{table}
\begin{table}[]
	\centering\scriptsize\caption{Same as Table \ref{T1} but for \textit{ec} and physical conditions listed below.  } \label{T8}
	\renewcommand{\arraystretch}{1.45}
	\begin{tabular}{cc|cc|c|cc|c}
		\hline
		\multicolumn{2}{}{}&\multicolumn{5}{c}{$\rho$ = 4.32(+07),~~~~~~~~~~~~$T_9$ = 3.26,~~~~~~~~~~~~$Y_e$ = 0.485}&\multicolumn{1}{}{} \\
		\hline
		\multicolumn{2}{}{}        & \multicolumn{2}{c}{Mass Fraction} & \multicolumn{1}{c}{$\lambda^{ec}$}      & \multicolumn{2}{c}{$\dot{Y}^{ec}_e$}  &            \\
		A  & Symbol & This work   & Nab21   & Nab21 & This work & Nab21 & Rank (Nab21) \\
		\hline
		56 & Ni & 1.53(-01) & 1.56(-01) & 8.43(-03) & 2.31(-05) & 2.35(-05) & 1  \\
		57 & Ni & 1.88(-02) & 1.77(-02) & 3.87(-02) & 1.28(-05) & 1.20(-05) & 2  \\
		55 & Co & 1.46(-02) & 1.52(-02) & 3.44(-02) & 9.11(-06) & 9.52(-06) & 3  \\
		58 & Ni & 2.97(-01) & 2.59(-01) & 3.54(-04) & 1.81(-06) & 1.58(-06) & 4  \\
		54 & Fe & 5.10(-01) & 5.45(-01) & 4.39(-05) & 4.14(-07) & 4.43(-07) & 5  \\
		53 & Fe & 3.54(-04) & 4.09(-04) & 4.10(-02) & 2.74(-07) & 3.16(-07) & 6  \\
		56 & Co & 5.91(-04) & 5.73(-04) & 1.79(-02) & 1.89(-07) & 1.84(-07) & 8  \\
		52 & Fe & 8.88(-04) & 1.11(-03) & 1.04(-02) & 1.78(-07) & 2.22(-07) & 7  \\
		59 & Cu & 1.58(-05) & 1.33(-05) & 3.30(-01) & 8.85(-08) & 7.45(-08) & 9  \\
		55 & Fe & 1.49(-03) & 1.47(-03) & 7.05(-04) & 1.90(-08) & 1.89(-08) & 10 \\
		53 & Mn & 2.01(-04) & 2.18(-04) & 2.08(-03) & 7.87(-09) & 8.55(-09) & 12 \\
		51 & Mn & 3.63(-05) & 4.59(-05) & 9.91(-03) & 7.06(-09) & 8.93(-09) & 11 \\
		59 & Ni & 2.69(-04) & 2.18(-04) & 1.48(-03) & 6.76(-09) & 5.47(-09) & 13 \\
		57 & Co & 6.80(-04) & 6.12(-04) & 4.63(-04) & 5.53(-09) & 4.97(-09) & 14 \\
		54 & Co & 7.03(-07) & 7.93(-07) & 2.73(-01) & 3.55(-09) & 4.01(-09) & 16 \\
		48 & Cr & 6.44(-06) & 9.56(-06) & 2.43(-02) & 3.26(-09) & 4.84(-09) & 15 \\
		52 & Mn & 1.13(-05) & 1.32(-05) & 1.46(-02) & 3.18(-09) & 3.72(-09) & 18 \\
		48 & V  & 5.87(-09) & 8.09(-09) & 2.31(-01) & 2.82(-09) & 3.89(-09) & 17 \\
		50 & Cr & 6.73(-04) & 8.58(-04) & 8.61(-05) & 1.16(-09) & 1.48(-09) & 19 \\
		49 & Cr & 1.94(-06) & 2.67(-06) & 2.63(-02) & 1.04(-09) & 1.43(-09) & 20 \\
		62 & Zn & 3.62(-06) & 2.51(-06) & 1.38(-02) & 8.06(-10) & 5.59(-10) & 22 \\
		58 & Cu & 9.52(-08) & 8.64(-08) & 4.32(-01) & 7.08(-10) & 6.43(-10) & 21 \\
		60 & Zn & 1.80(-07) & 1.45(-07) & 1.88(-01) & 5.64(-10) & 4.55(-10) & 23 \\
		60 & Cu & 4.66(-07) & 3.64(-07) & 7.03(-02) & 5.46(-10) & 4.26(-10) & 24 \\
		51 & Cr & 1.69(-06) & 2.00(-06) & 6.10(-03) & 2.02(-10) & 2.39(-10) & 25 \\
		61 & Zn & 1.19(-08) & 8.94(-09) & 2.82(-01) & 5.53(-11) & 4.14(-11) & 29 \\
		46 & V  & 6.94(-11) & 1.11(-10) & 3.58(-01) & 5.40(-11) & 8.66(-11) & 26 \\
		61 & Cu & 2.57(-06) & 1.86(-06) & 8.69(-04) & 3.66(-11) & 2.65(-11) & 32 \\
		40 & Ca & 1.83(-05) & 3.53(-05) & 7.24(-05) & 3.31(-11) & 6.40(-11) & 27 \\
		57 & Cu & 7.32(-10) & 7.17(-10) & 2.24(-00) & 2.87(-11) & 2.82(-11) & 31\\
		\hline
	\end{tabular}
\end{table}
\begin{table}[]
	\centering\scriptsize\caption{List of top 50 most relevant \textit{ec} and \textit{bd} nuclei averaged over the entire stellar trajectory for $0.500 > {Y_e} > 0.400$. Nuclei marked with asterisk are new entries (appeared due to Coulomb corrections) not to be found in the list compiled by Nab21.  } \label{T9}
	\renewcommand{\arraystretch}{1.75}
	\renewcommand{\tabcolsep}{0.1cm}
	\begin{tabular}{cccccc|cccccc}
		\hline
		\multicolumn{6}{c}{\textit{ec} nuclei}  &
		\multicolumn{6}{c}{\textit{bd} nulcei} \\
		  & $\mathring{R}_{p}$ & Rank &    & $\mathring{R}_{p}$ &  Rank &  & $\mathring{R}_{p}$ &  Rank &  & $\mathring{R}_{p}$ &  Rank \\
		Nuclei  & This work & Nab21 &  Nuclei  & This work &  Nab21 &Nuclei  &This work &  Nab21 &Nuclei  & This work&  Nab21 \\
		\hline
		$^{56}$Mn            & 1.21(+01)            & 1                    & $^{56}$Fe            & 4.53(-02)            & 21                   & $^{53}$Mn            & 6.03(-02)            & 7                    & $^{69}$Cu            & 5.11(-03)            & 12                   \\
		$^{67}$Cu            & 9.75(-01)            & 3                    & $^{64}$Ni            & 4.41(-02)            & 36                   & $^{55}$Mn            & 5.19(-02)            & 10                   & $^{54}$Cr            & 4.98(-03)            & 38                   \\
		$^{60}$Co            & 7.04(-01)            & 4                    & $^{75}$Ge$^\ast$            & 3.99(-02)            & 68                   & $^{67}$Ni            & 4.51(-02)            & 1                    & $^{51}$Sc            & 4.62(-03)            & 15                   \\
		$^{66}$Cu            & 4.16(-01)            & 7                    & $^{57}$Ni            & 3.97(-02)            & 25                   & $^{63}$Co            & 3.56(-02)            & 4                    & $^{58}$Cr            & 4.32(-03)            & 11                   \\
		$^{52}$V             & 3.67(-01)            & 2                    & $^{86}$Kr$^\ast$            & 3.97(-02)            & 111                  & $^{57}$Fe            & 3.19(-02)            & 16                   & $^{53}$Ti            & 4.29(-03)            & 22                   \\
		$^{53}$Mn            & 2.18(-01)            & 5                    & $^{55}$Co            & 3.95(-02)            & 23                   & $^{49}$Sc            & 3.00(-02)            & 2                    & $^{63}$Fe            & 3.95(-03)            & 26                   \\
		$^{59}$Co            & 1.79(-01)            & 10                   & $^{72}$Ga$^\ast$            & 3.89(-02)            & 65                   & $^{65}$Co            & 2.95(-02)            & 3                    & $^{51}$Cr            & 3.59(-03)            & 46                   \\
		$^{65}$Cu            & 1.70(-01)            & 22                   & $^{75}$Ga            & 3.88(-02)            & 24                   & $^{57}$Mn            & 2.46(-02)            & 20                   & $^{53}$V             & 3.44(-03)            & 40                   \\
		$^{55}$Fe            & 1.63(-01)            & 9                    & $^{86}$Br$^\ast$            & 3.80(-02)            & 78                   & $^{57}$Co            & 2.38(-02)            & 21                   & $^{59}$Co$^\ast$            & 3.42(-03)            & 52                   \\
		$^{85}$Br$^\ast$            & 1.43(-01)            & 51                   & $^{57}$Fe            & 3.77(-02)            & 34                   & $^{55}$Cr            & 2.34(-02)            & 18                   & $^{54}$Mn            & 3.26(-03)            & 50                   \\
		$^{61}$Ni            & 1.01(-01)            & 13                   & $^{77}$Ga            & 3.68(-02)            & 12                   & $^{55}$Fe            & 1.79(-02)            & 24                   & $^{75}$Ga            & 3.23(-03)            & 35                   \\
		$^{79}$Ge            & 9.90(-02)            & 11                   & $^{58}$Ni            & 3.53(-02)            & 33                   & $^{64}$Co            & 1.64(-02)            & 8                    & $^{71}$Cu            & 3.16(-03)            & 29                   \\
		$^{83}$Se            & 9.70(-02)            & 38                   & $^{57}$Co            & 3.48(-02)            & 31                   & $^{59}$Mn            & 1.49(-02)            & 6                    & $^{49}$Ca            & 3.15(-03)            & 9                    \\
		$^{78}$Ge            & 9.67(-02)            & 14                   & $^{53}$Cr            & 3.44(-02)            & 39                   & $^{52}$V             & 1.40(-02)            & 23                   & $^{67}$Co            & 2.56(-03)            & 25                   \\
		$^{49}$Sc            & 8.02(-02)            & 6                    & $^{63}$Ni$^\ast$            & 3.39(-02)            & 60                   & $^{51}$Ti            & 1.32(-02)            & 14                   & $^{58}$Mn            & 2.37(-03)            & 41                   \\
		$^{68}$Cu            & 6.22(-02)            & 19                   & $^{69}$Cu            & 3.14(-02)            & 26                   & $^{58}$Fe            & 1.28(-02)            & 28                   & $^{85}$Se$^\ast$            & 2.36(-03)            & 68                   \\
		$^{82}$Se$^\ast$            & 6.12(-02)            & 67                   & $^{54}$Cr            & 3.02(-02)            & 27                   & $^{61}$Fe            & 1.01(-02)            & 13                   & $^{60}$Mn            & 2.19(-03)            & 31                   \\
		$^{56}$Ni            & 6.08(-02)            & 18                   & $^{58}$Co            & 2.96(-02)            & 37                   & $^{62}$Fe            & 7.30(-03)            & 19                   & $^{54}$V             & 2.14(-03)            & 33                   \\
		$^{87}$Kr$^\ast$            & 6.08(-02)            & 91                   & $^{65}$Ni            & 2.70(-02)            & 42                   & $^{50}$Sc            & 7.01(-03)            & 5                    & $^{65}$Ni$^\ast$            & 1.77(-03)            & 53                   \\
		$^{83}$As            & 5.78(-02)            & 15                   & $^{67}$Ni            & 2.60(-02)            & 17                   & $^{53}$Cr            & 6.58(-03)            & 34                   & $^{59}$Fe$^\ast$            & 1.77(-03)            & 65                   \\
		$^{82}$As            & 5.62(-02)            & 20                   & $^{55}$Mn            & 2.50(-02)            & 47                   & $^{77}$Ga            & 6.36(-03)            & 32                   & $^{87}$Se$^\ast$            & 1.67(-03)            & 59                   \\
		$^{77}$Ge            & 5.40(-02)            & 35                   & $^{81}$Se$^\ast$            & 2.36(-02)            & 95                   & $^{56}$Fe            & 6.26(-03)            & 39                   & $^{70}$Cu            & 1.63(-03)            & 37                   \\
		$^{64}$Cu            & 5.07(-02)            & 30                   & $^{61}$Co$^\ast$            & 2.32(-02)            & 61                   & $^{57}$Cr            & 5.75(-03)            & 17                   & $^{79}$Ga            & 1.58(-03)            & 36                   \\
		$^{73}$Ga            & 4.82(-02)            & 49                   & $^{51}$Ti            & 2.28(-02)            & 16                   & $^{61}$Co            & 5.57(-03)            & 45                   & $^{84}$As$^\ast$            & 1.49(-03)            & 56                   \\
		$^{63}$Cu            & 4.62(-02)            & 32                   & $^{85}$Se$^\ast$            & 2.23(-02)            & 64                   & $^{66}$Co            & 5.47(-03)            & 27                   & $^{68}$Ni            & 1.45(-03)            & 47                   \\
		\hline
	\end{tabular}
\end{table}
\end{document}